\documentclass[english,useAMS,usenatbib]{mn2e}
\usepackage[T1]{fontenc}
\usepackage[latin1]{inputenc}
\usepackage{array}
\usepackage{textcomp}
\usepackage{graphicx}
\usepackage{amssymb}
\usepackage{array}
\providecommand{\tabularnewline}{\\}
\usepackage{booktabs}
\usepackage{babel}
\usepackage{times}

\title{Resolved Imaging of the HD191089 Debris Disc}

\author[Churcher, Wyatt \&  Smith]{Laura Churcher$^1$\thanks{e-mail:ljc51@ast.cam.ac.uk}, Mark Wyatt$^1$, Rachel Smith$^2$,\\$^1$Institue of Astronomy, University of Cambridge, Madingley Road, Cambridge, UK, CB3 0HA\\ $^2$Astrophysics Group, Keele University, Staffordshire, UK, ST5 5BG}

\begin{document}
\maketitle
\date{Accepted: Submitted:}

\pagerange{\pageref{firstpage}--\pageref{lastpage}} \pubyear{2010}
\label{firstpage} 
\begin{abstract}
Two thirds of the F star members of the 12 Myr old $\beta$ Pictoris
Moving Group (BPMG) show significant excess emission in the mid-infrared,
several million years after the expected dispersal of the protoplanetary
disc. Theoretical models of planet formation suggest that this peak
in the mid-infrared emission could be due to the formation of Pluto-sized
bodies in the disc, which ignite the collisional cascade and enhance
the production of small dust. Here we present resolved mid-infrared
imaging of the disc of HD191089 (F5V in the BPMG) and consider its
implications for the state of planet formation in this system. HD191089
was observed at $18.3\umu m$ using T-ReCS on Gemini South and the
images were compared to models of the disc to constrain the radial
distribution of the dust. The emission observed at $18.3\umu m$ is
shown to be significantly extended beyond the PSF at a position angle
of 80$^{\circ}$. This is the first time dust emission has been resolved
around HD191089. Modelling indicates that the emission arises from
a dust belt from 28-90 AU, inclined at $35^{\circ}$ from edge on
with very little emission from the inner 28AU of the disc, indicating
the presence of an inner cavity. The steep slope of the inner edge is more
consistent with truncation by a planet than with ongoing stirring.
A tentative brightness asymmetry $\frac{F_{W}}{F_{E}}=0.80\pm0.12$
(1.8$\sigma$) between the two sides of the disc could be evidence
for perturbations from a massive body on an eccentric orbit in the
system. 
\end{abstract}

\section{Introduction}

The majority ($\geqslant$80 per cent) of stars are born with protoplanetary
discs (\citealp{2001ApJ...553L.153H}) which are massive and contain
large amounts of gas, with an estimated gas to dust ratio of 100:1
(\citealp{2005ApJ...631.1134A}). These discs have a lifetime of $\sim$6Myr
(\citealp{2001ApJ...553L.153H}) and at ages of a few Myr undergo
a transition, the nature of which is still unclear (\citealp{2010ApJ...708.1107M}).
Some protoplanetary discs have inner holes (\citealp{2010ApJs..186..111L})
but the systems retain an optically thick outer disc and are often called transition discs to signify that this is a phase that discs go through. At later times ($\sim$10Myr) the remaining optically
thick material in the disc has also disappeared, leaving optically
thin debris discs.

 Debris discs surround $\sim$15 per cent of main sequence
stars (see \citet{2008ARA&A..46..339W} for a recent review). The dust particles
in debris discs are generally small, and so are subject to radiation
forces such as radiation pressure and Poynting-Robertson (P-R) drag
from the central star (\citealp{1993prpl.conf.1253B}) which remove
grains from the system on timescales much shorter than the stellar
lifetime so the dust must be continuously replenished. The collisional
destruction or sublimation of larger bodies (planetesimals) can create
such dust particles. Effectively, the small dust can be used as a
tracer for the presence of bodies many hundreds of kilometers in size
that are otherwise undetectable. Inner cavities have also been
observed in debris discs, either directly through images or indirectly through an absence of hot dust. The holes imply a lack of planetesimals in the inner region, as rapid collisional processing means that none of the dust created in the outer disc reaches the  inner region (\citealp{2006ApJ...639.1153W}).

Since gas giants must form in the gas-rich protoplanetary disc then it is possible that the planetary system is fully formed by the end of the transition phase. Imaging of debris discs at ages of $\sim$ 10Myr shows evidence that planet formation in these systems
has already finished and formed Jupiter mass planets. The existence of such planets was inferred from debris disc observations (\citealp{2006MNRAS.372L..14Q,2005ApJ...620..492W}) and subsequently confirmed through direct imaging of planets in systems such as the $\sim$8M$_{jup}$
planet imaged around $\beta$ Pic, a 10Myr old A star with a debris
disc (\citealp{2009A&A...493L..21L}) and a <3M$_{jup}$ planet imaged around Fomalhaut (\citealp{2008Sci...322.1345K,2009ApJ...693..734C}).

However models indicate that the growth of solid objects in a debris disc can continue for several 100 Myr ({\citealp{2004ApJ...602L.133K}). In such self-stirred models a bright ring appears in an extended planetisimal disc where Pluto sized bodies have just formed. These bodies stir the planetesimal belt into collisional destruction and increase dust production, causing the disc to brighten at this location. There is some evidence in favour of this; observations of a
peak in the brightness of debris discs around A stars at ages of 5
to 10 Myr (\citealp{2008ApJ...672..558C}) can be explained in this context but a more detailed examination indicates that these models still require an inner hole to reproduce the observations. The inner region could have been cleared by a planet that is then stirring the outer regions of the disc (\citealp{2010MNRAS.tmp..493K}).

 Resolved imaging of young
debris discs is therefore important in discriminating between discs
being perturbed by a fully formed planet, such as in the pericentre
glow model applied to HR4796 (\citealp{1999ApJ...527..918W}) and
discs in which planet formation may still be ongoing such as $\eta$
Tel (\citealp{2009A&A...493..299S}) where the disc is modelled using
the delayed stirring models of \citet{2004ApJ...602L.133K}.

Many young stars are too distant to study their discs in
detail, but young moving groups in this age range, such as the BPMG
(12Myr) provide a relatively nearby, young population in which to
assess the origin and diversity of debris disc emission at this age (\citealp{2004ARA&A..42..685Z}).

HD191089 (HIP 99273) is an F5V star at a distance of 54pc (\citealp{1997AA...323L..49P})
that was identified as a candidate debris disc by \citet{1998ApJ...497..330M}
based on IRAS photometry. The age of this source is somewhat uncertain
and estimates in the literature vary widely. Isochrone fitting suggests
older ages, with \citet{2004A&A...418..989N} giving an age of 3
Gyr and \citet{2006ApJS..166..351C} finding 1.6 Gyr. However, when
considering evidence of youth such as X-Ray emission and Lithium abundance
data amongst other techniques, the age indicated is significantly
younger; \citet{2004ARA&A..42..685Z} find 100Myr. However, \citet{2006ApJ...644..525M}
suggest that the source is a possible member of the $\beta$ Pictoris
Moving Group (BPMG) due to HD191089's proper motion and high rotation
rate, giving it an age of 12 Myr. This membership appears to be confirmed
from further proper motion studies by \citet{2009AJ....137.3632L}.
The BPMG is a rich source of debris discs with $\sim$23 per cent (\citealp{2008ApJ...681.1484R})
of members harbouring a disc. Here, we adopt an age of 12Myr when
interpreting this source.

Using the Spitzer fluxes for this source (\citealp{2008ApJ...677..630H}) and other sun-like stars, an analysis similar to that presented in \citet{2010arXiv1004.0644S} for A star debris discs showed that HD191089 was the most favourable target for imaging debris discs with deep mid-IR observations.

In \S2 we present 18.3$\umu$m imaging of HD191089 taken with the
T-ReCS instrument on Gemini South. In \S3 we analyse these observations
and show that we have resolved the debris disc. In \S4 we confront
these observations with models to determine the disc parameters and
in \S5 discuss the implications of the inferred structure for the
status of planet formation in this system, both in the context of
the Pericentre Glow model (\citealp{1999ApJ...527..918W}) and the
self-stirred model  (\citealp{2004ApJ...602L.133K}).

\section{Observations}

These observations were taken under proposal GS-2009A-28 using T-ReCS
on Gemini South with filter Qa ($\lambda_{c}=18.3\umu m$, $\Delta\lambda=1.51\umu m$).
The pixel scale of the T-ReCS instrument is 0''.09 and the total
field of view of is 29''$\times$22''. These observations were taken
in parallel chop-nod mode with a chop throw of 15'' and chop position
angle of $55^{\circ}$ (East of North). The data were taken over three
nights (9th June 2009, 12th July 2009, 13th July 2009). The total
on source time was 7200s. Observations of standard star HD189831 (spectral
type K5III) listed in \citet{1999AJ....117.1864C}, were made to
calibrate the photometry and also to monitor the PSF of the observations.
The order of the observations and the on source integration times
are shown in Table \ref{tab:Observations-taken-under}.

\begin{table*}
\caption{\label{tab:Observations-taken-under}Observations taken under proposal
GS-2009A-Q-28 in order. Note that the
on source time is half of the total integration time listed in the
table. Fluxes are for a 1'' radius aperture centered on the star.
The group indicates the standard-science-standard observing pattern
used. One standard was used in groups 2 and 3 due to an omitted calibration
observation. Each science observation consists of 4 sub integrations.}

\begin{tabular}{|c|c|c|c|c|c|c|}
\hline 
Date  & Object  & Group & Name  & Filter  & Integration Time (s)  & Calibrated Flux (mJy)\tabularnewline
\hline 
09/06/09  & HD189831  & 1  & Std1 & QA  & 120  & 3441\tabularnewline
09/06/09  & HD191089  & 1  &  Im1 & QA  & 1440  & $158\pm21$\tabularnewline
09/06/09  & HD189831  & 1  & Std2 & QA  & 120  & 3441\tabularnewline
\noalign{\vskip\doublerulesep} 12/07/09  & HD189831  & 2 & Std3  & QA  & 1440  & 3441\tabularnewline
12/07/09  & HD191089  & 2  & Im2 & QA  & 1440  & $167\pm$22\tabularnewline
12/07/09  & HD189831  & 2/3  &  Std4 & QA  & 120  & 3441\tabularnewline
12/07/09  & HD191089  & 3  & Im3 & QA  & 1440  & $166\pm23$\tabularnewline
12/07/09  & HD189831  & 3  &  Std5 &QA  & 120  & 3441\tabularnewline
12/07/09  & HD189831  & 4  &  Std6 & QA  & 120  & 3441\tabularnewline
12/07/09  & HD191089  & 4  & Im4 & QA  & 1440  & $145\pm$24\tabularnewline
12/07/09  & HD189831  & 4  & Std7 & QA  & 120  & 3441\tabularnewline
\noalign{\vskip\doublerulesep} 13/07/09  & HD189831  & 5  & Std8 & QA  & 120  & 3441\tabularnewline
13/07/09  & HD191890  & 5  & Im5 & QA  & 1440  & $174\pm$31\tabularnewline
13/07/09  & HD189831  & 5  & Std9 & QA  & 120  & 3441\tabularnewline
\hline
\end{tabular}
\end{table*}

These data for HD191089 were taken in sets of four 360s long integrations
with 6 complete nod cycles (ABBA). The data were reduced using custom
routines described in \citet{2008A&A...485..897S}. The data reduction
involved determining a gain map for each observation using the mean
values of each frame to construct a map of pixel responsitivity. The
on-source pixels were masked during this process, making this equivalent
to a sky flat field frame. A DC offset was then determined by calculating
the mean pixel values in every row and every column, again masking
pixels where there was source emission present. This was then subtracted
from the final image to ensure a flat background. Pixels which showed
high or low gain in comparison with the median response throughout
the observation were masked off. In order to avoid errors in co-adding
the data which could arise from misalignment of the images,   we fitted a Gaussian with a sub-pixel centroid to accurately determine the centre}
of the image and so the position of the star. The re-binning was done
using bilinear interpolation across the array.

The standard star was observed at a similar airmass to HD191089. The
calibration levels were compared with the airmass for the standard
star observations and no correlation was found, so no extinction correction
was applied to the calibration factors.

In order to calibrate the images an average of the calibration factors
determined from standard star observations was used. The standard
star, HD189831 ($F_{18}=3.44$Jy from \citealp{1999AJ....117.1864C,2001A&A...367..297A})
was observed twice per group of observations (See Table \ref{tab:Observations-taken-under}
for detail of Observations Groups). A coadd of the two standards from
each group was used to calibrate the observations of HD191089 in the
same group. The average calibration error for the five groups of observations
was 6 per cent. The groups had calibration factors that varied by 3 per cent, 6 per cent,
12 per cent, 6 per cent and 4 per cent. These values were used to calculated a calibration
error for the fluxes. These centred, flux calibrated images from each
group were then used to produce final images of HD191089 and the standard
star.

\section{Results}

\begin{figure*}
\caption{\label{fig:Obs}Left to Right: The final co-added images of the standard
star (left) and HD 191089 (centre) and a residual emission image shown
after subtracting an azimuthally symmetric PSF created from the standard
star image scaled to the expected photospheric emission from HD191089
(44mJy) from the science image (centre). Note that the distinct ellipticity
seen in the image of HD191089 is not seen in the standard star. Contours
on the left and central plots are at 25, 50 and 75 per cent of the peak (solid
lines) and 10 per cent of the peak (dashed line). The contours on the residual
emission image are at 25 per cent, 50 per cent and 75 per cent of the peak. Orientation
of the images is North up, East left. The peak value for HD191089
in QA is 344 mJy/arcsec$^{2}$ and the peak value for the standard
is 27648 mJy/arcsec$^{2}$. The peak value for the residuals is 55mJy/arcsec$^{2}$.
The residual image has been smoothed by convolving with a Gaussian
with a width of 4 pixels.The colour scale is linear from 0 to the
maximum pixel value in each image.}\vspace{10mm}

\includegraphics[width=6cm,height=6cm, keepaspectratio, bb=150 50 520 400 ]{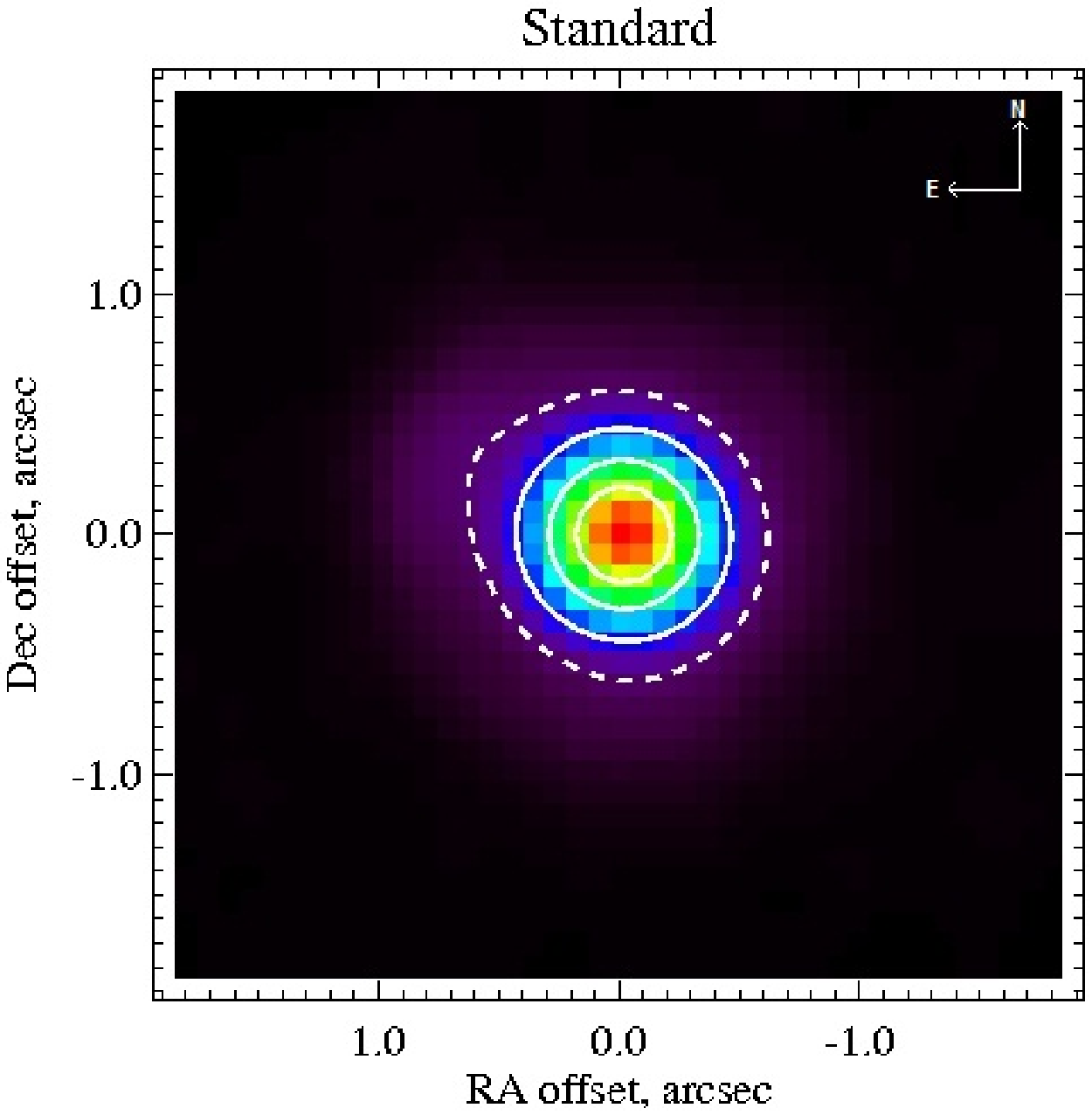}\includegraphics[width=7.5cm,height=7.5cm,keepaspectratio, bb=50 50 520 400]{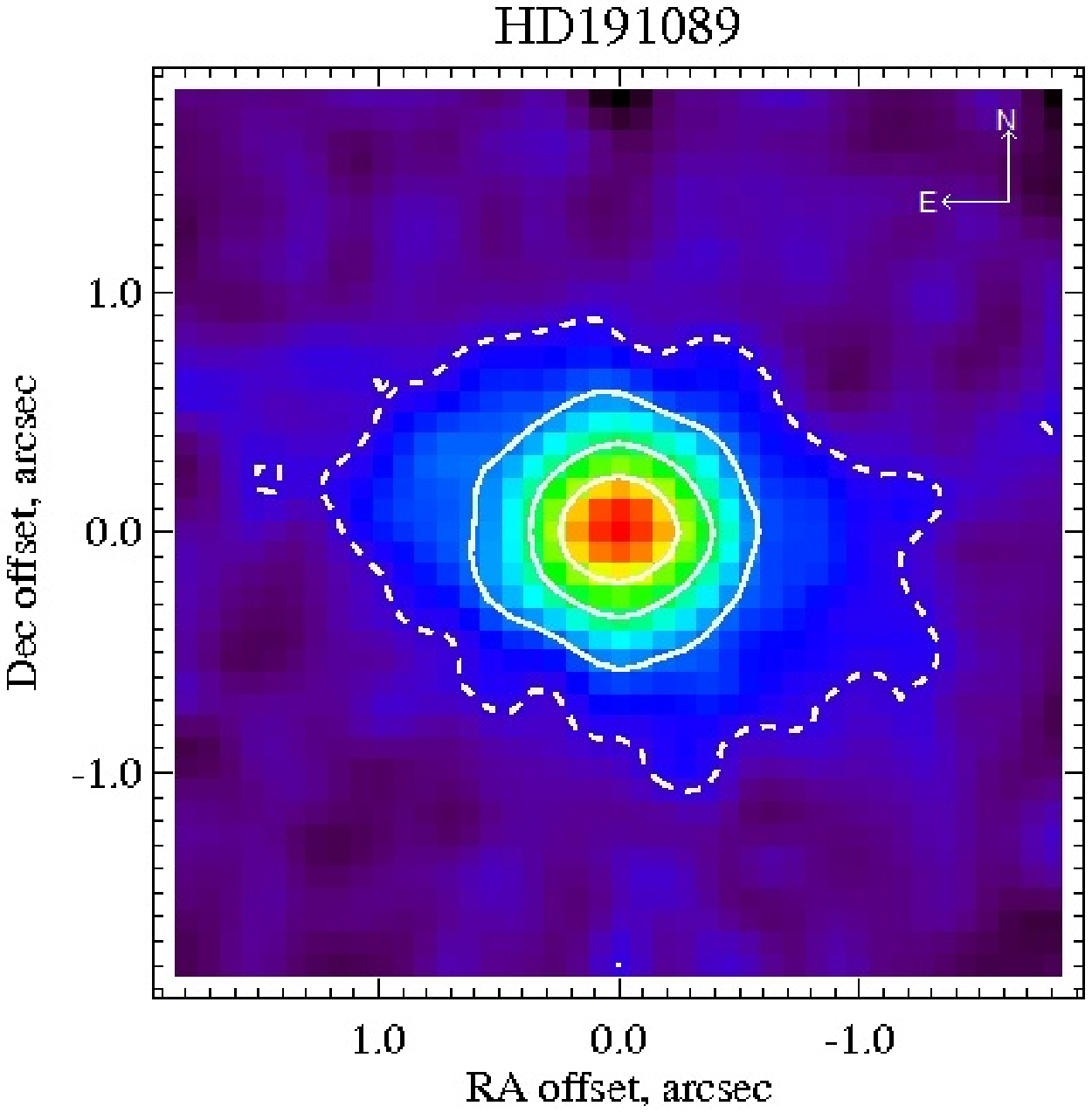}\includegraphics[width=4.5cm,height=3.5cm, bb=10  30 400 400 ]{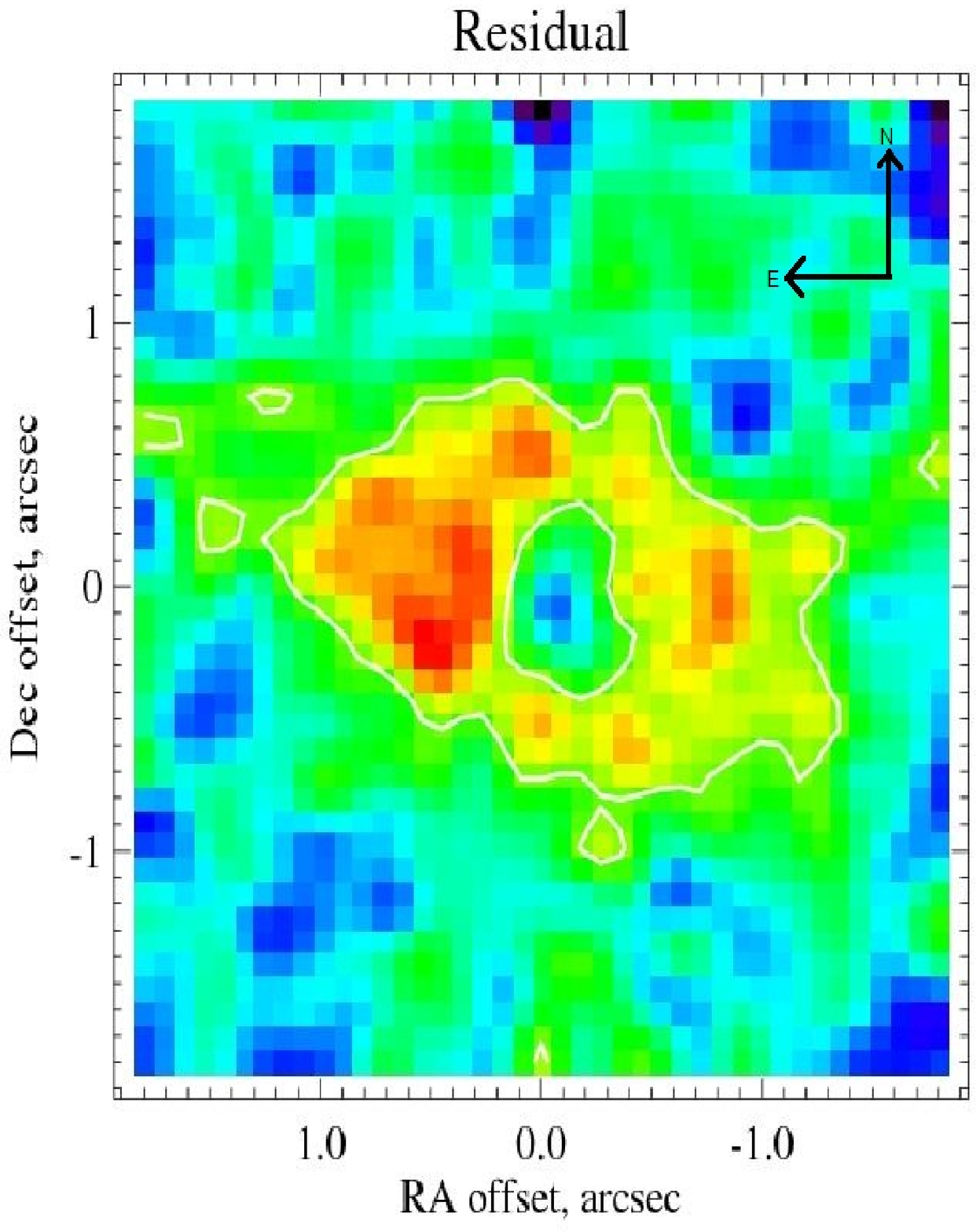} \vspace{1.5mm}
\end{figure*}

The final co-added images for HD191089 and the standard star HD189831
for PSF reference are shown in Figure \ref{fig:Obs}. The solid line
contours shown in the images are 25 per cent, 50 per cent and 75 per cent of the peak for
both HD191089 and the standard. The dashed contour represents 10 per cent
of the peak emission. The peak value for HD191089 in the final coadded
QA image is 344mJy/arcsec$^{2}$ and the peak value for the standard
is 27648 mJy/arcsec$^{2}$.

Photometry was performed using a 1''.0 radius circular aperture centred
on the star. The stellar flux in the Qa filter is expected to be 44mJy
from the K band flux (see Section 4.2). Our photometry yields a flux
of 163$\pm$22 mJy including both calibration and photometric errors
for the final co-added image of HD191089 shown in Figure \ref{fig:Obs}
with an S/N of 38 (not including calibration errors). Thus the photometry
gives an excess of 119$\pm$22 mJy in a 1'' radius aperture centred
on the star. The IRS spectrum of \citealp{2006ApJS..166..351C} after
subtraction of the stellar flux gives an excess of 115$\pm$3mJy.
The T-ReCs photometry presented here therefore agrees with the IRS
results. It does not appear that the IRS spectrum includes
any emission centred on the star outside the 1'' aperture used here
for photometry, as might happen in the larger IRS beam (extracted
along a 3.7'' slit). The total error on the photometric measurements
in Table \ref{tab:Observations-taken-under} consists of both the
calibration error and the statistical noise. The statistical noise
was determined using an annulus with an inner radius of 3'' and an
outer radius of 4'' centred on the star, resulting in an error averaged
over all the groups of 1.1mJy/arcsec$^{2}$. 

There is clear ellipticity in the image of HD191089 (centre of Figure
\ref{fig:Obs}) and a 2D Gaussian fit to HD191089 gives the ellipticity
of the image at the 50 per cent contour as $0.104\pm$0.015 with the major axis at position angle
of $80^{\circ}\pm10^{\circ}$ East of North. The same ellipticity
is seen on all three nights. In comparison the ellipticity of the
standard was $0.024\pm0.007$ with the major axis at a position angle
of $55^{\circ}\pm2^{\circ}$ East of North. As all the observations
were performed with no on-sky rotation and chop-nod performed at $55^{\circ}$
East of North, the ellipticity in the PSF could be due to chop smearing,
but the extension seen in the HD19089 images is at a very different
position angle and so is unlikely to be an artifact of the chopping
process.

The right image in Figure \ref{fig:Obs} is the residual image of
HD191089 after the subtraction of the standard star image scaled to
the flux expected from the photosphere of HD191089 (44mJy) followed
by smoothing by a Gaussian of width 4 pixels. Scaling the subtraction
to the expected stellar flux means that we are not removing any of
the disc emission. We also tried subtracting the standard star image
scaled to the peak of the HD191089 image, to see if there was any
additional unresolved flux present at the peak, and found that this
results in a subtraction of 49mJy, indicating that there is very little
unresolved disc flux contributing to the peak. A ring is seen in the
residual image, consistent with a disc that is at some intermediate
inclination between edge on and face on. The position angle of the
peaks (and hence the major axis of the ring) was found by finding
the angle between the peaks in the residual image and was found to be
at $80^{\circ}\pm10^{\circ}$ East of North, consistent with the orientation
of the ellipticity found for the science observations before subtraction
of the PSF.

\begin{figure}
\caption{\label{fig:FWHMs}The profile of the line cuts through the total images
at $80^{\circ}$ E of N (top) and $170^{\circ}$E of N (bottom). The
median FWHM measurements taken over all sub-integrations from Gaussian
fits to the profiles at the two angle are indicated for reference.
The averaged profiles from all the line cuts for both HD191089 and
the standard star are also shown. The profile for HD191089 (dashed
line) at $80^{\circ}$ is much broader in the wings than that of the
standard star (solid line). There is also some evidence of a less
pronounced extension at $170^{\circ}$ as seen in the lower image.
Note that multiple intensity profiles for the science target (diamonds)
and the standard (crosses) are shown here. These are taken for each
group as indicated in table 1.}

\includegraphics[width=7cm,height=5cm]{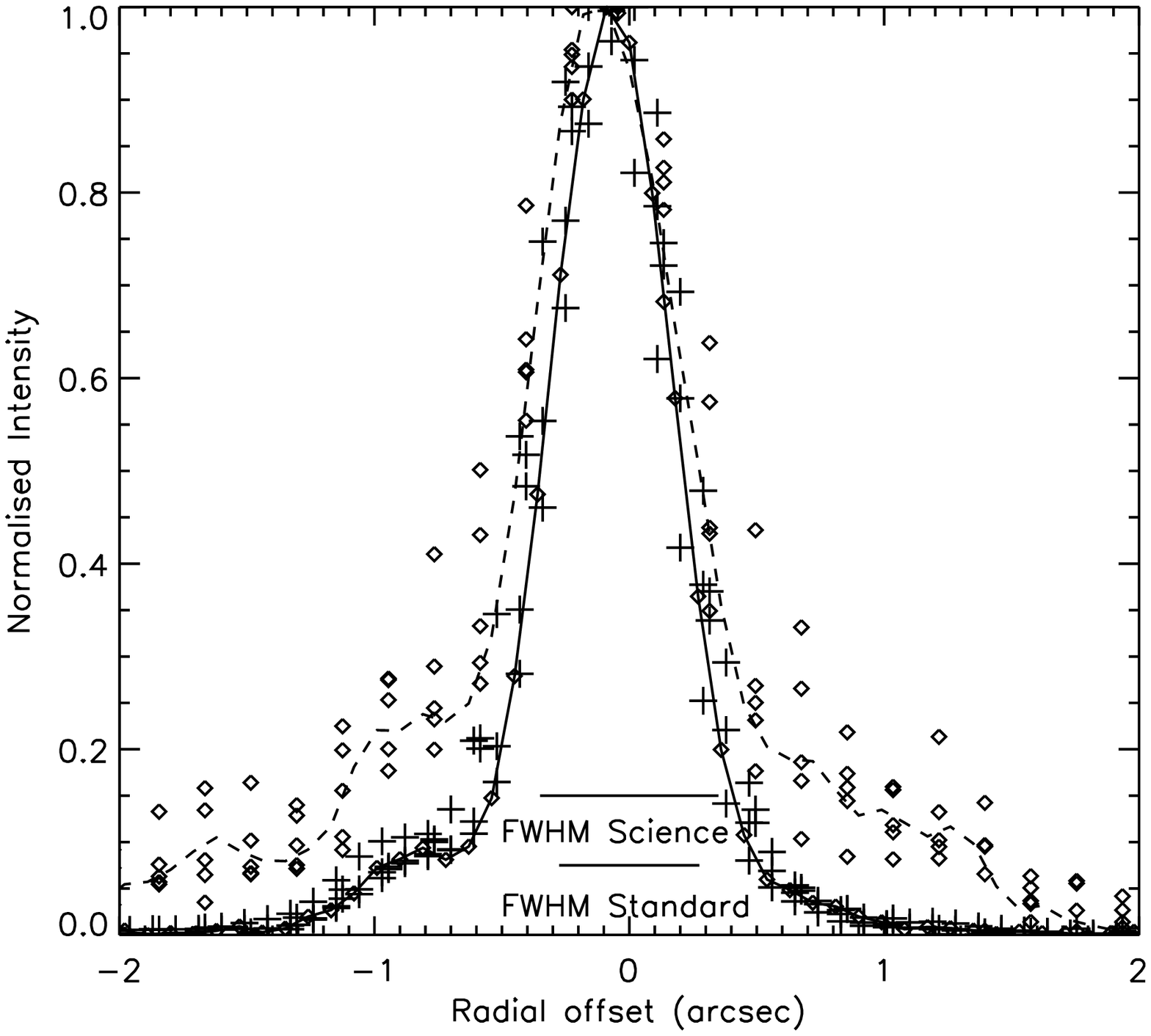}

\includegraphics[width=7cm,height=5cm]{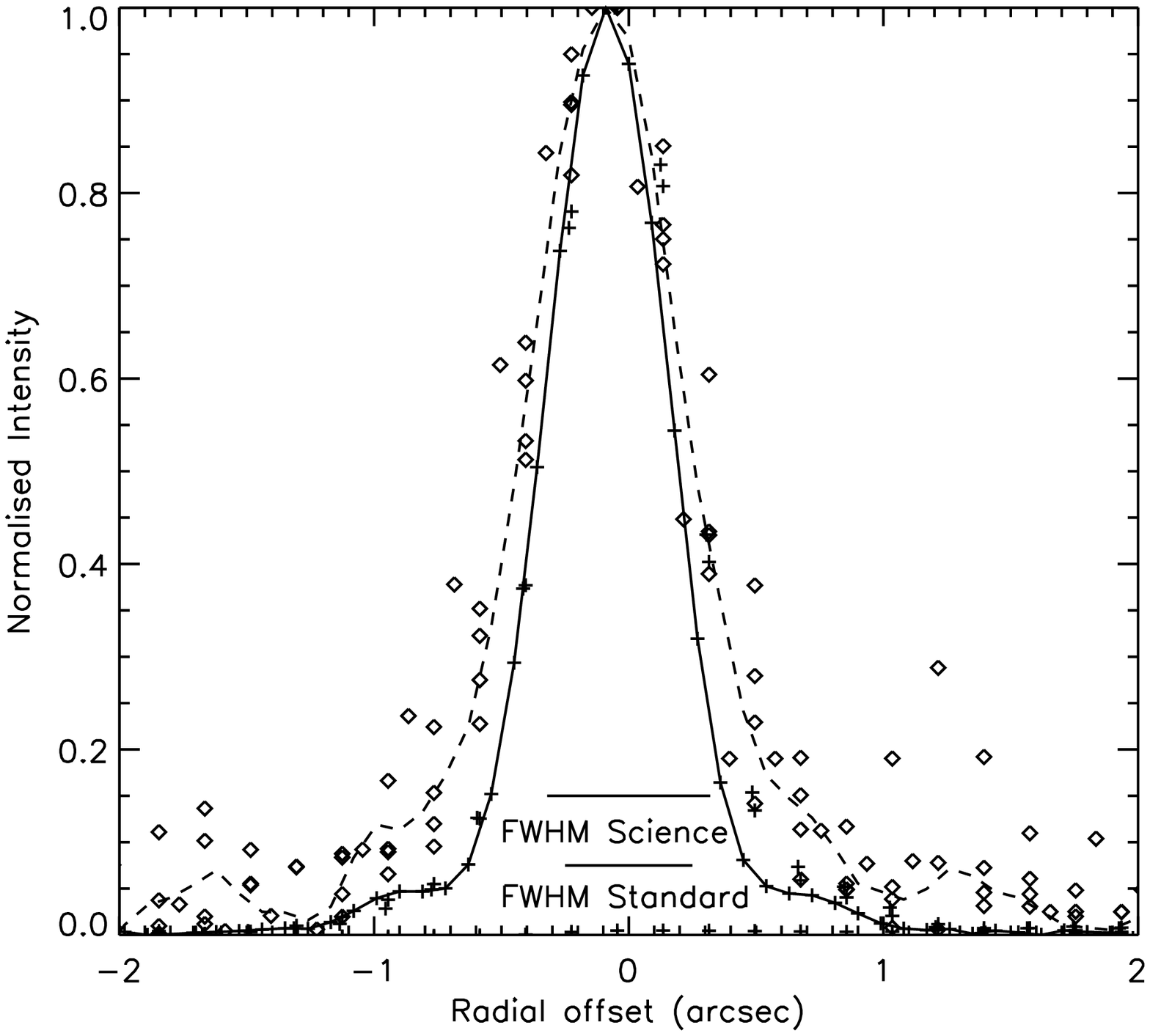} 
\end{figure}

In order to assess the robustness of this putative disc, and to ensure
that it is not an artifact of the PSF, we examined the temporal variability
of the FWHM of the science data and assessed the significance of the
extension. We also looked for any dependence of the FWHM on time or
airmass. We took linecuts at position angles corresponding to the
major and minor axes of the fitted ellipticity in the science images
(PA of $80^{\circ}$ and $170^{\circ}$) for each group of HD191089
observations (see Table \ref{tab:Observations-taken-under} for observation
groups) and for PSF standards from each observational group. The linecuts
were smoothed by a Gaussian of width 3 pixels and were then fitted
with a Gaussian profile to determine the FWHM. No dependence on airmass
or time was found, and the median FWHM values and standard errors
were found to be: $8.81\pm$0.2 pixels at $80^{\circ}$for the standard,
$9.95\pm$0.3 pixels at $80^{\circ}$ for HD191089, $8.10\pm$0.1 pixels at $170^{\circ}$
for the standard, and $9.46\pm$0.4 pixels for HD191089 at $170^{\circ}$.
There is naturally a large degree of scatter in the intensity profiles,  due to variations in observing conditions,
but HD191089 shows extension when compared to the standard
at $80^{\circ}$ and there is also some evidence of extension along
the minor axis of the image at $170^{\circ}$. This is shown graphically
in Figure \ref{fig:FWHMs} which shows the linecuts through each integration
(of 2880s, see table \ref{tab:Observations-taken-under}) of HD191089
(diamonds) and corresponding standard star integrations (plus signs).
HD191089 also has broader wings than the standard star image at $80^{\circ}$
.

The residual image of HD191089 shown in Figure \ref{fig:Obs} also
shows tentative evidence for a brightness asymmetry between the two
sides of the disc. The ratio of the flux in a 0.5'' radius aperture,
centred on the E peak (found by fitting a Gaussian at sub-pixel resolution)
and the flux in an identical aperture centred at the same radial location
on the W side was $\frac{F_{W}}{F_{E}}=0.80$. This aperture size
was chosen to maximise the asymmetry observed. An azimuthally symmetric
PSF was used for the subtraction shown in Fig. \ref{fig:Obs}, so
this apparent brightness difference is not simply due to non-axisymmetric
features in the PSF. This was also checked by looking for any systematic
azimuthal variations in the PSF, and by redoing the subtraction using
each of the 10 standards to quantify the effect of PSF variability
on the measured asymmetry.  It was found that variability in the PSF FWHM and asymmetries change
the flux ratio by $\pm$0.02, or $<0.2\sigma$ significance.  The final stacked PSF image used had a FWHM of 8.82 pixels at 80$^\circ$ and 8.10 pixels at 170$^\circ$.

The significance of this putative asymmetry was assessed using a Monte
Carlo method. We created 10 000 axisymmetric disc models (see Section
4.1), convolved these models with the axisymmetric PSF and added Poissonian
noise consistent with the observed levels in the images ($\sim$0.1mJy/pixel).
This does not account for the effects of any spatially correlated
noise, such as coherent variation in the PSF over the course of the
observations, but when we examined the PSF from each group of observations
(see Table \ref{tab:Observations-taken-under}) we found no systematic
asymmetries. This modelling provides a first guess at the chance of
observing such an asymmetry simply due to uncertainty from Poissonian
noise in the images. The average observed asymmetry between the two
sides of the image in these noisy asymmetric models was 12 per cent. The
brightness asymmetry observed is then significant at the $1.8\sigma$
level. 

\section{Modelling \label{sec:Modelling}}

\subsection{Axisymmetric Disc Model \label{sub:Axisymmetric-Disc-Model}}

To determine constraints on the radial distribution of the emission
seen in the resolved 18$\umu m$ image we first considered the simplest
possible model of the disc structure. This model is composed of a
single axisymmetric disc component, defined by four free parameters:
disc radius (r), width (dr), inclination (i) and surface density distribution
$\Sigma$, which is assumed to have the form $\Sigma\propto r^{\gamma}$.
The disc opening angle (which sets the disc height) is assumed to
be 5$^{\circ}$ but this parameter is unconstrained by the modelling
process. The flux from annuli in the disc at different radii was determined
assuming a grain composition and size distribution that were constrained
using the emission spectrum as outlined in section 4.2. Since the SED fitting
process fixes the total flux from the disc model to be $\sim$115mJy this 
did not have a significant impact on the derived spatial distribution.

\begin{figure*}
\caption{\label{fig:Model Im}  The best fitting axisymmetric disc model for
the 18$\umu m$ emission (see section 4.1). The
model residual image (left0 can be compared directly
with the observed residuals (Figure \ref{fig:Obs}, right).  The difference between the observed residuals and the model residuals (right) indicates an over subtraction near the
 disc ansae and a slight under subtraction at the peak. See
section \ref{sub:Inner-hole-and} for details.}

\includegraphics[width=6.5cm, height=6.5cm]{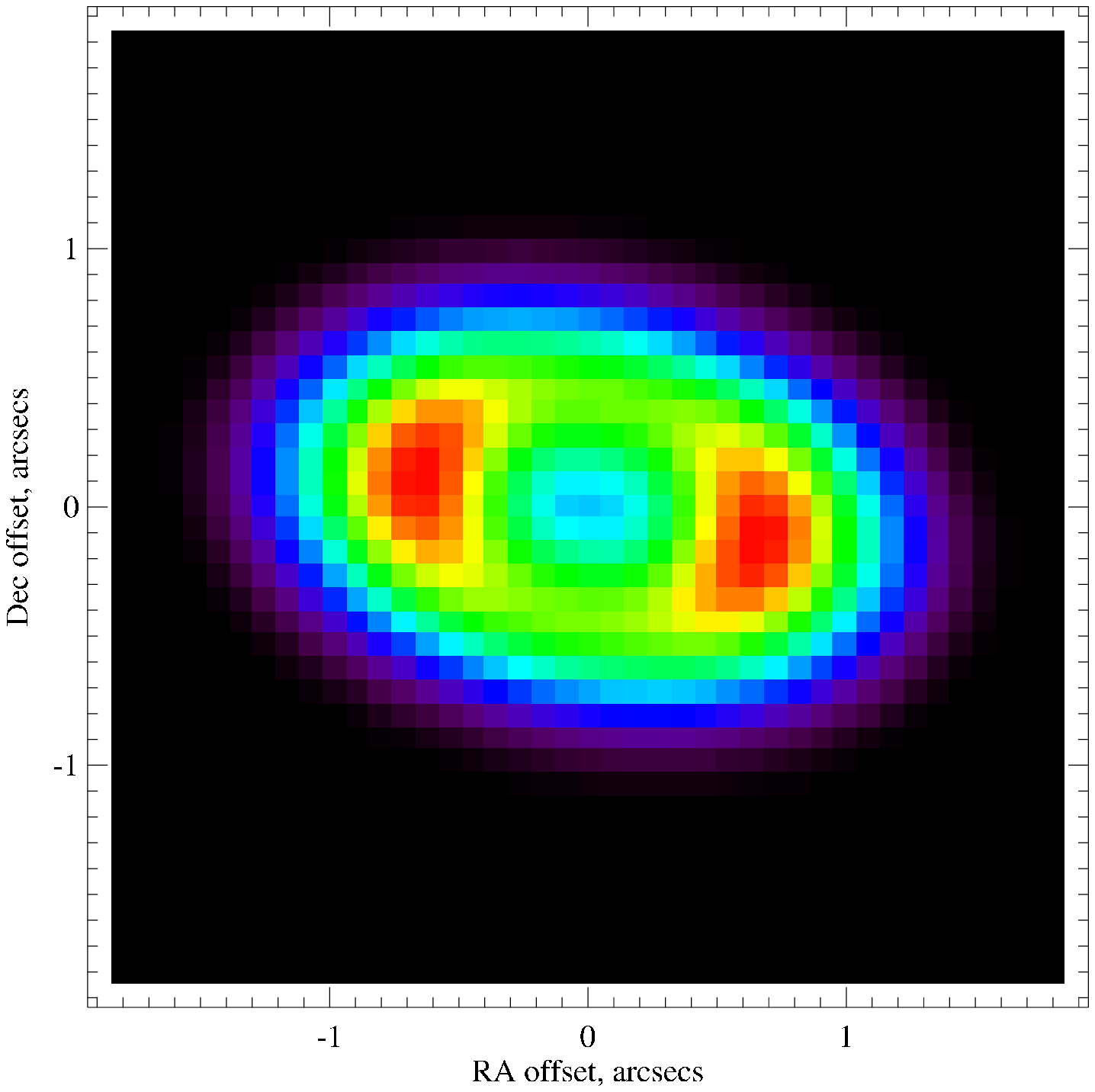}\ \ \ \ \   \includegraphics[width=6.5cm,height=6.5cm]{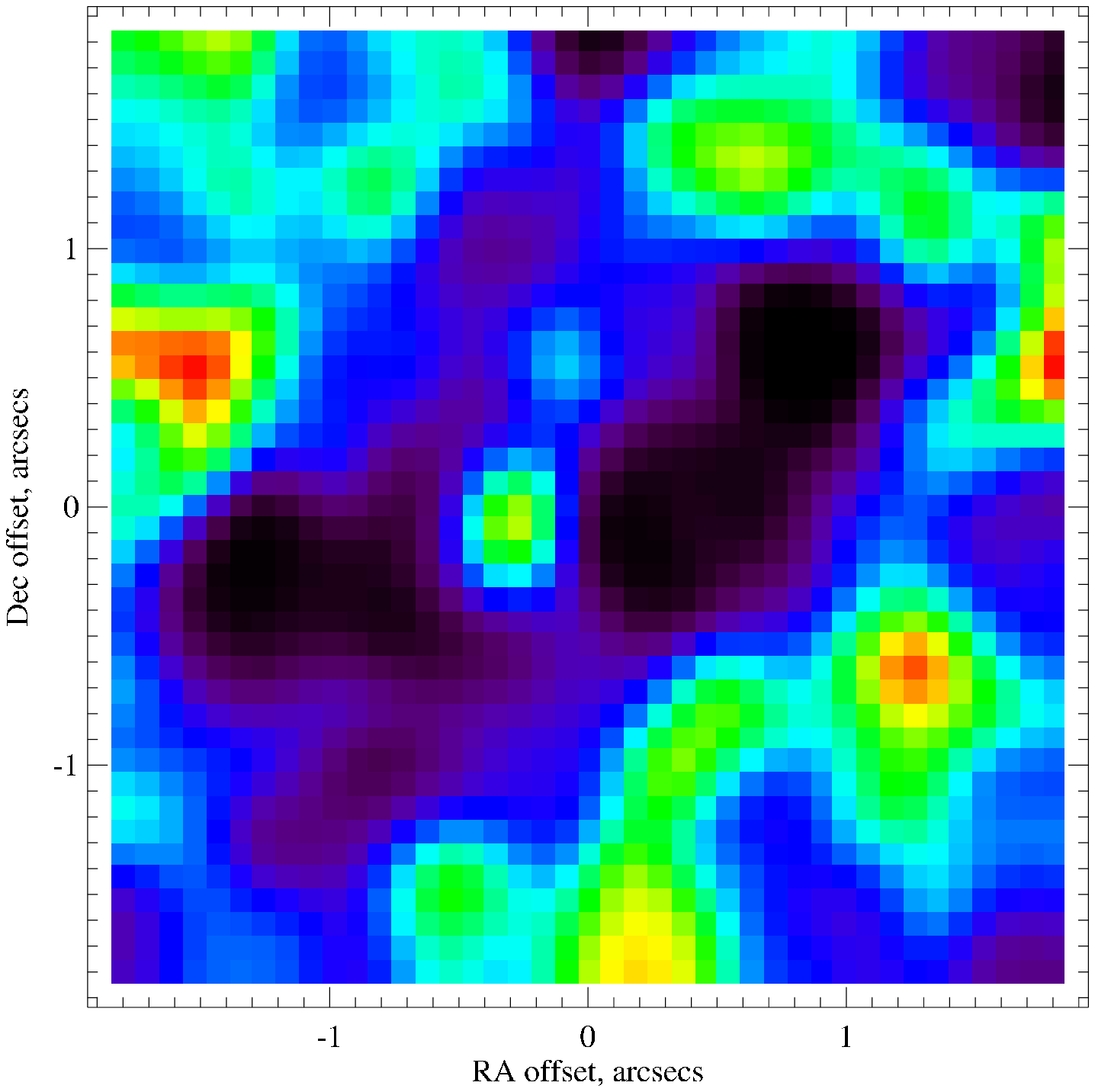} 
\end{figure*}

In order to constrain the dust location, as well as the width of the
disc, the inclination and the surface density profile, a grid of models
was run. The model images were convolved with an azimuthally symmetric
PSF and compared with the observed residuals using  the images as well as the linecuts both in the direction
of extension (PA $80^{\circ}$) and perpendicular to the extension
(PA $170^{\circ}$). The use of both linecuts allowed us to constrain
the inclination simultaneously with the radial morphology. The reduced
chi-squared of the fit ($\chi_{r}^{2}=\frac{(obs-mod)^{2}}{\nu}=1$
where $\nu$ is the number of free parameters and 1 represents a good
fit) to each of these three pieces of observational data were calculated
and were then combined linearly with equal weight to come to a final
mean reduced chi-squared ($\chi_{r_{combined}}^{2}=\frac{\chi_{rline_{80}}^{2}+\chi_{rline_{170}}^{2}+\chi_{rres}^{2}}{3}$)
which was then minimised across the grid of models. The ranges of
model parameters tested were $R_{in}$:10 to 50AU (1 AU intervals),
$R_{out}$: 50 to 100AU (5 AU intervals), inclination: 10$^{\circ}$
to 60$^{\circ}$ ($5{}^{\circ}$ intervals, where 0$^{\circ}$ is
edge on), surface density index $\gamma$: -0.5 to -3.0 (0.5 intervals).
The values for gamma were chosen to cover possibilities such as the
surface density distribution expected from grains being blown out
of the system by radiation pressure ($\gamma$=-1.0), and that of
the Minimum Mass Solar Nebular (MMSN) ($\gamma$=-1.5). 
The best fit model was found to be a ring at 28$\pm1$ to 90$\pm5$AU, with an inclination
of 35$\pm$5$^{\circ}$ from edge on, and a surface density profile
$\gamma$=-1.5$\pm0.5$. This best fit model had $\chi_{r_{combined}}^{2}=1.30$
with $\chi_{rline_{80}}^{2}$=1.34, $\chi_{rline_{170}}^{2}$=1.40,
$\chi_{rres}^{2}=1.19$. The best fit model residual image is shown on the left of Figure \ref{fig:Model Im}
and can be compared directly with the observed disc (Fig \ref{fig:Obs}
right) which has the same colour scale.

PSF variation can have an important impact on the observational residuals,
and we used an azimuthally symmetric average PSF. This was motivated
by the lack of effects of temporal and spatial variation in the PSF
(see Section 3). The modelling process for the axisymmetric disc was
repeated using the PSF with the largest FWHM (Std 8)  and the smallest
FWHM (Std 6) to find that PSF variation had no effect on the modelling
results within the quoted errors.

\subsection{SED Models \label{sub:Combining-Imaging-and}}

\begin{figure*}
\caption{\label{fig:Spectral-Energy-Distribution} Spectral Energy Distribution
(SED) of HD191089. The photosphere of HD191089 is fitted with a Kurucz
model profile ($M_{star}=1.4M_{\odot}$, $L_{star}=3.2L_{\odot}$,
$T_{star}=6440K$) scaled to the 2MASS K band flux and
shown with a solid line. Observations at $>5\umu m$ are shown after
the subtraction of the photospheric contributions: the IRS spectrum
of \citealp{2006ApJS..166..351C} (solid line from $5-30\umu m$),
MIPS fluxes (\citealp{2009ApJS..181..197C}) (triangles) and IRAS
colour-corrected excesses (asterixes) and Gemini TReCs fluxes in 0.5''
radius aperture (square), and sub-mm/mm upper limits (crosses) from
SHARC and CARMA. The excess measurements are first fitted with a 99K
modified blackbody (dashed line), for comparison to a realistic grain fit (solid
line) described in section 4.2}.

\includegraphics[width=9cm,height=8cm]{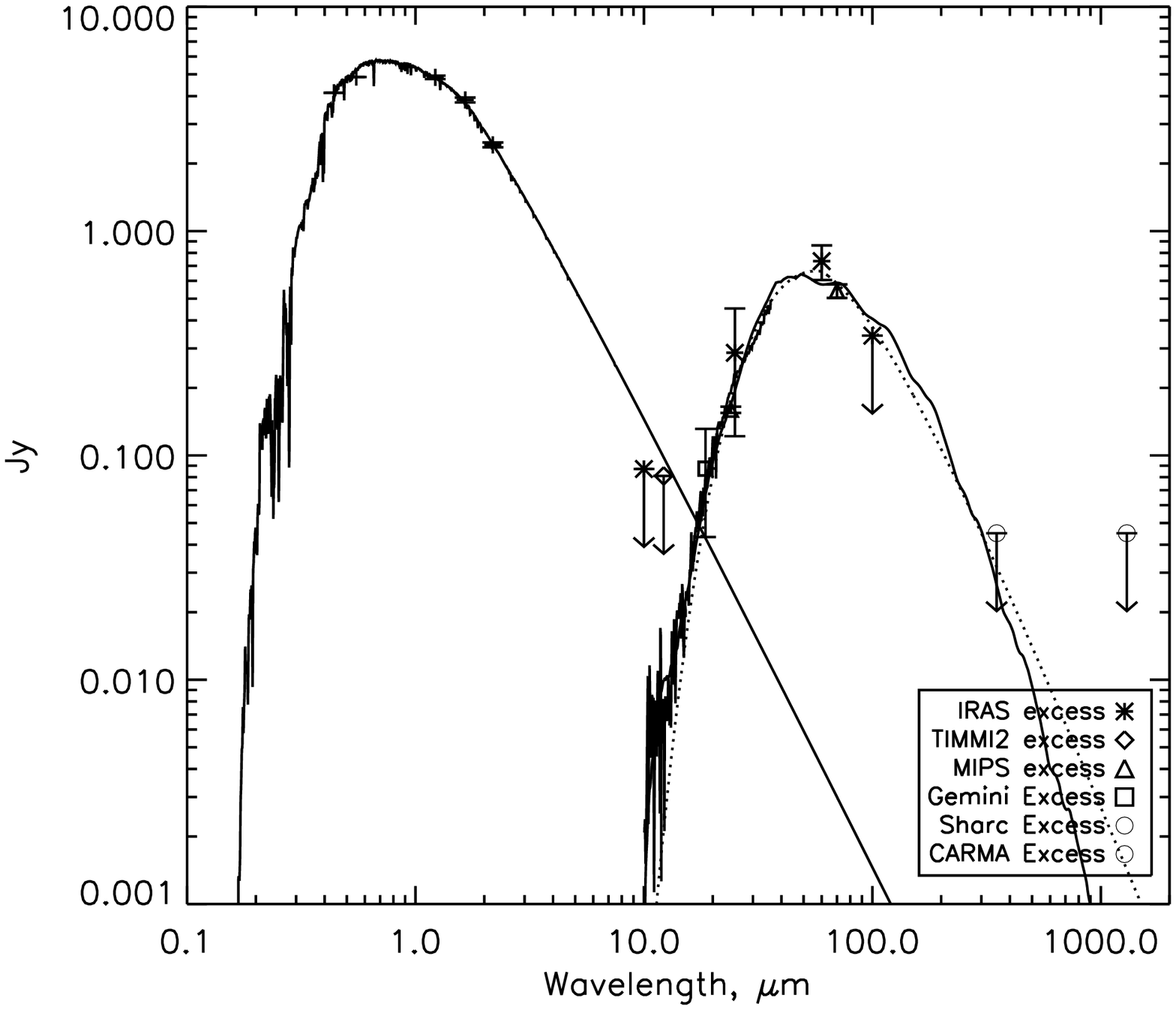}\includegraphics[width=9cm,height=8cm,keepaspectratio]{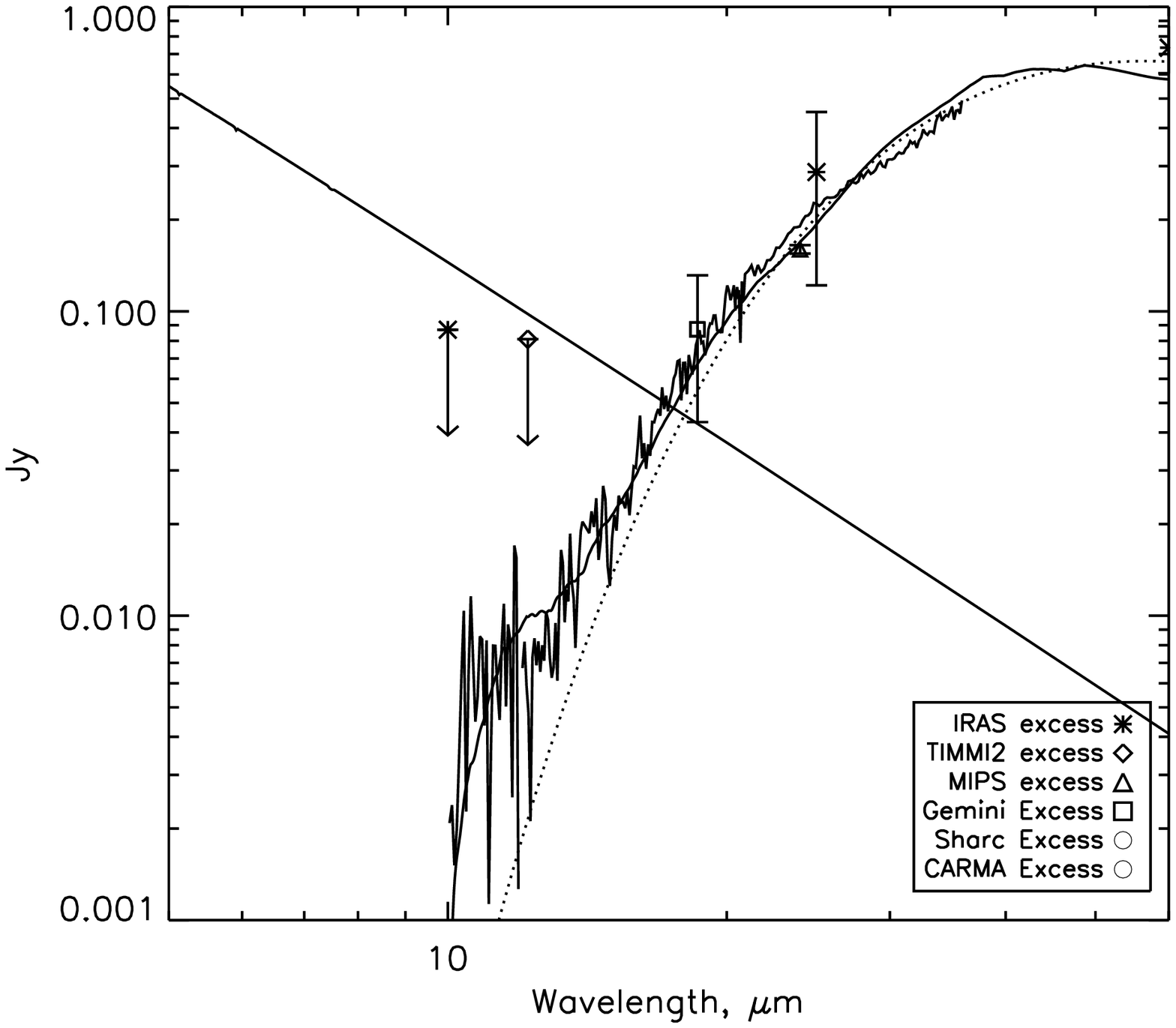} 
\end{figure*}

The emission spectrum of HD191089 is shown in Figure \ref{fig:Spectral-Energy-Distribution}.
The stellar photospheric emission has been modelled using a Kurucz
model atmosphere with a temperature of 6440K (\citealp{1979ApJS...40....1K}),
fitted to the 2MASS K band photometry implying a stellar luminosity
of $3.2L_{\odot}$ at the inferred distance of 53.5 pc. Photospheric
flux contributions have been subtracted from the fluxes shown at $>5\umu$m
which thus represent the excess emission from the disc. This spectrum
is approximately that of a blackbody fit with a temperature of $99\pm9$K
and a fractional luminosity ($f=\frac{L_{dust}}{L_{*}}$) of $1.5\times10^{-3}$
(\citealp{2006ApJS..166..351C}).

To calculate the emission from grains in the model they were assumed 
to have a size distribution with $n(D)\propto D^{-3.5}$,  where D is the grain diameter, which is
the standard solution for a theoretical collisional cascade (see \citealp{1969JGR....74.2531D}),
that is truncated at a minimum and maximum
 grain size. For the assumed
composition a grid of models was calculated with a range of minimum
sizes 0.5-10$D_{bl}$ where $D_{bl}$ is the largest grain size removed
from the system by radiation pressure for which $\beta$=0.5 where
$\beta=\frac{F_{rad}}{F_{grav}}$, the ratio between radiation and
gravitational forces (\citealp{1979Icar...40....1B}) which depends
on stellar properties, grain composition and porosity. The maximum
size was fixed at 1cm since larger grains have a negligible contribution
to flux from a model with this size distribution.  The grains were
assumed to have a silicate core (amorphous olivine) and an accreted mantle of organic refractories  produced by UV photoprocessing of ice.
(as used in \citealp{1997A&A...323..566L,1999A&A...348..557A}). A
range of compositions was also tried, with amorphous silicate fractions
varying from 0 per cent to 90 per cent by volume and with porosities (i.e. vacuum
fraction of grain by volume) from 0 per cent to 95 per cent.  Dielectric constants
were calculated from tabulated laboratory values (\citealp{1997A&A...323..566L,1999A&A...348..557A}) using Maxwell-Garnett effective medium theory.  The
optical properties of the grains were calculated using Mie theory, Rayleigh-Gans theory
and Geometric Optics in the appropriate size regimes (\citealp{1983asls.book.....B}).

This resulted in a grid of 360 models over which $\chi^{2}$ was minimised
by fitting to the spectrum and MIPS fluxes ($24\umu m$ and 70$\umu m$).
The fit to the IRS spectrum was calculated using $\chi_{IRS}^{2}=\frac{(F_{obs}-F_{mod})^{2}}{N}$
where $F_{obs}$ is the observed flux, $F_{mod}$ is the model flux
and N is the number of points in the observed IRS spectrum. This normalises
$\chi_{IRS}^{2}$ so that when calculating the final $\chi^{2}$ the
fit to the spectrum was given equal weight to the fits to the 24$\umu$m
and 70$\umu$m MIPS fluxes. The best fit model with a reduced $\chi^{2}$
of 1.14 has a minimum grain size that is coincident with the blow-out
grain size (1.46$\umu m$ for the composition used) and a composition
of amorphous silicate fraction of 10 per cent by volume, and a porosity of
60 per cent  with the rest of the grain composed of organic refractory materials,
with no ices present. The fractional luminosity is $1.4\times10^{-3}$.
As there are no obvious spectral features in the SED, the main constraint
on the composition and size distribution comes from allowing the appropriate
range of temperatures to be present given the constraint that the
dust is in the region 28-90 AU; blackbody grains would have
to be at a radius of 15 AU to achieve the observed temperature. Although
our model provides a consistent fit to both image and SED, it is not
expected that the composition has been uniquely constrained by this
process.This dust model was then used as an input to the model grid and there was no change to the best fitting model of the 18$\umu$m images.

The total mass in the collisional cascade, with the assumed size distribution
of $n(D)\propto{D^{-3.5}}$ scales as $M_{tot}\propto \sqrt{D_{max}}$.  Thus for $D_{max}=1$km the total
mass would be 13$M_{\oplus}$ two orders of magnitude higher than the current mass of the Kuiper Belt ( \citealp{2004AJ....128.1364B}), but note that we have no information
on the maximum size of objects in this system.

\subsection{Emptiness of the Inner Hole}

It is possible to fit all excesses in the spectrum with a model that
only extends from 28AU-90AU. There is no requirement either from the image or from the spectrum
for emission at <28AU. The observed 18$\umu$m image of HD191089 shows a resolved
belt of dust from 28-90 AU, with a cleared inner region extending
to a radius of $\sim$28AU. We compared the photometry of the observations in the inner 28AU (0.5'' radius aperture)
and then repeated the measurement with the same aperture on the models, where an unresolved flux component could be added at the position of the star. Excluding calibration uncertainties, the measured flux density within a 0.5" radius aperture on the 18$\umu$m image was 48$\pm$4mJy of flux, with a photospheric contribution of 33mJy (75 per cent of total photospheric flux, measured in the same aperture using a scaled PSF).   The same measurement on the model with no stellar or unresolved flux component added gives 16mJy.Therefore the maximum unresolved component that can be added to the model that does not exceed a 3$\sigma$ upper limit of the observations is 12mJy. However these measurements do not account for the 12 per cent calibration uncertainty in the photometry or the error in photospheric flux due to uncertainties in the spectral fitting to the star, which is $\pm$1.2mJy (on total photospheric flux). Since a change in calibration would also result in a corresponding change in the mode fluxl (to fit the peaks at the correct level), it is instructive to consider what would have been derived with a revised calibration factor and stellar flux. In the extreme situation that the calibration was  36 per cent higher  (i.e. a 3$\sigma$ upper limit on the calibration factor) and the stellar flux was lower by 3.2mJy (3$\sigma$ lower limit), the flux measured in the previously described apertures would have been 65$\pm$5mJy with a corresponding model flux of 22mJy. Thus the excess would have been13$\pm$5mJy and so gives a maximum unresolved component of <28mJy for a 3$\sigma$ deviation in both calibration and statistical uncertainty.

\subsection{Pericentre Glow Model \label{sub:Pericentre-Glow-Model}}

\begin{figure*}
\caption{\label{fig:PGMOD}The best fitting disc from the pericentre glow model
which is scaled and shown in the same way as the HD191089 residuals
(Figure \ref{fig:Obs}) for comparison. The disc parameters are the
same as those for the best-fitting axisymmetric disc shown in Figure
\ref{fig:Model Im} (28-90AU, i=$35^{\circ}$, $\Sigma{\propto}r^{\gamma}$) Right:
linecut through the observed and model images at a PA of 80$^{\circ}$
(Observed: solid line, Model: dotted line)}

\includegraphics[width=6cm,height=6cm]{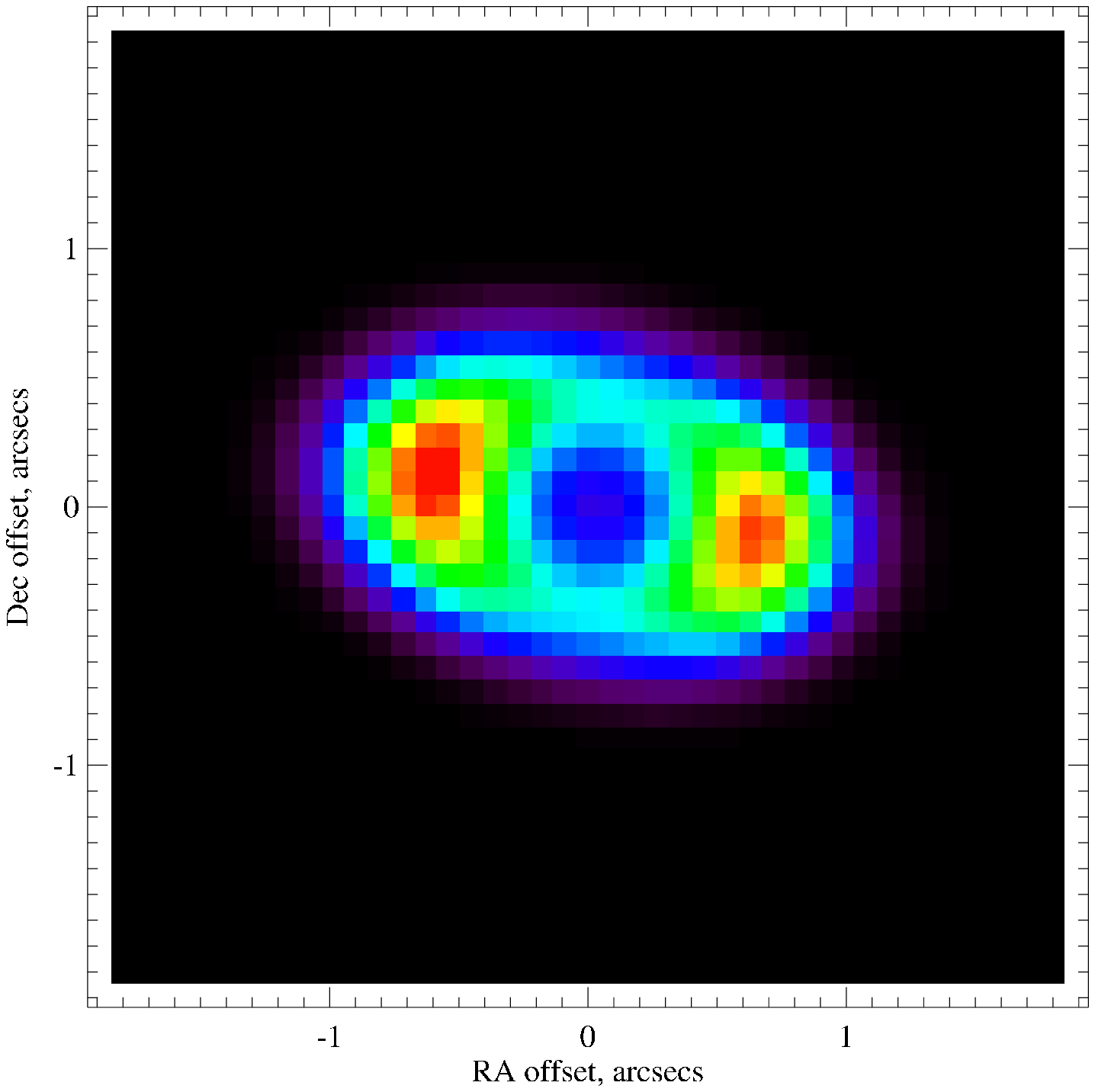}\includegraphics[width=8cm,height=6cm]{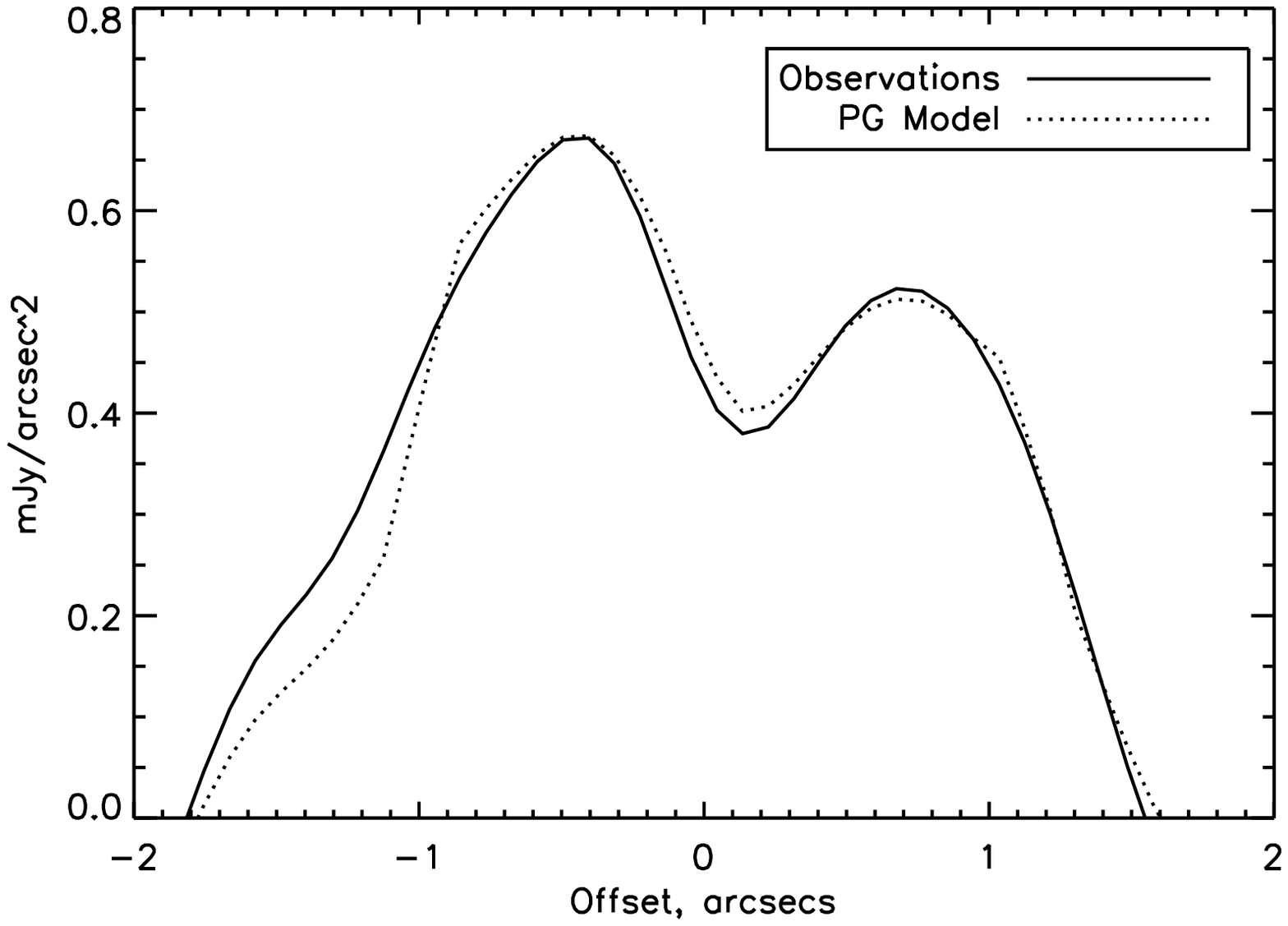} 
\end{figure*}

The brightness asymmetry seen in Figure \ref{fig:Obs} is reminiscent
of that seen in the disc of HR4796 (\citealp{2000ApJ...530..329T}).
The 6 per cent asymmetry of HR4796 was interpreted as a consequence of pericentre
glow, where the secular gravitational perturbations of a planet on
an eccentric orbit imposes a forced eccentricity on the planetesimal
belt, causing the forced pericentre side of the disc to be closer
to the star and hence hotter and brighter (\citealp{1999ApJ...527..918W}).
Here we applied the same model to the HD191089 disc. The disc is defined
by the inner and outer semi-major axis of the disc, a$_{min}$ and
a$_{max}$ with a proper eccentricity e$_{p}$=0 (so a$_{min}$ and
a$_{max}$ are equivalent to r$_{min}$ and r$_{max}$ in the axisymmetric
model), the power law index for the distribution of semi-major axes $\gamma_{PG}=\gamma$+1 where $\gamma$
is the surface density index from the axisymmetric model, and a forced
eccentricity e$_{f}$. The best fit model has parameters a$_{min}$=28,
a$_{max}$=90, $\gamma_{PG}=2.5$ and e$_{f}$=0.12 and is shown
in Figure \ref{fig:PGMOD}.

This addition of an extra parameter (the forced eccentricity) improves
the $\chi_{r_{combined}}^{2}$ from 1.31 to 1.15, but in general adding
an extra parameter will improve the fit of any model. However, given
the tentative (1.8$\sigma$) nature of the asymmetry, is adding this
extra parameter to the model justified? We assess the justification
for this extra parameter through use of the Bayesian Information Criterion
(BIC):

\[
BIC=Nln(\chi_{r_{combined}}^{2})+klnN\]

where N is the number of data points, k is the number of free parameters
and $\chi_{r_{combined}}^{2}$is the minimum combined reduced $\chi^{2}$
for the model. The BIC considers the fit of the model but penalises
the model for extra parameters(see \citealp{2005ApJ...618..385W,2007MNRAS.377L..74L}). 
Two models were considered: the axisymmetric disc model(see section
\ref{sub:Axisymmetric-Disc-Model}) and the pericentre glow model
(see section \ref{sub:Pericentre-Glow-Model}). The BIC value for
the axisymmetric disc is 19.3 and for the pericentre glow model is
is 19.2. There is very little difference between the BIC values, indicating
that there is no preference between the two models.

\section{Discussion}

\subsection{\label{sub:Inner-hole-and}Implications for the Planetary System}

There are 3 different possible interpretations of the state of planet
formation in the HD191089 system based on the resolved image: \textit{i)
}The inner cavity and brightness asymmetry are due to the formation
of planets interior to the inner edge of the resolved disc (\ref{sub:Planet-in-the}),
\textit{ii) }The disc is in the process of forming Pluto-sized bodies
as in the delayed stirring models of \citet{2005AJ...130..269K}
(\ref{sub:Delayed-Stirring-Model})or \textit{iii) }The disc is a
remnant of the protoplanetary disc (\ref{sub:Protoplanetary-Disc-Remnant}).

\subsubsection{\label{sub:Planet-in-the}Planet in the Central Cavity}

The 18$\umu$m image of HD191089 shows a resolved belt of dust from 28-90 AU, inclined at 35$^\circ$ from edge on with a cleared inner region extending to a radius of $\sim$28AU. 
The brightness asymmetry seen in the HD191089 disc could be evidence
for a planet on an eccentric orbit (see Section \ref{sub:Pericentre-Glow-Model}).
The pericentre glow model only places constraints on the forced eccentricity
imposed on the debris disc by the planet's secular perturbations and
this cannot be used to constrain its mass and semi-major axis. The
only constraint is that the lower the mass of the planet the longer
the secular precession timescales. 

Although it is not known it seems
likely that should such a planet exist, the same planet that is causing the asymmetry is causing
the inner edge, as proposed by \citet{2006MNRAS.372L..14Q}.
The inner edge of the resolved dust belt at 28AU could therefore mark
the outer edge of the planetary system, especially if a giant planet
is sculpting the inner edge of the disc, as in the case of Fomalhaut
(\citealp{2008Sci...322.1345K,2009ApJ...693..734C}) which has a dust
belt with a well defined inner edge and a brightness asymmetry.

 The steepness of the inner edge of the disc also  suggests truncation by a planet. For a disc that extends
inward to $\sim$5AU, like the discs seen
in \citet{2004ApJ...602L.133K}, a surface density of $\Sigma\propto r^{6}$
from 5-28AU is required to satisfy the observed 3$\sigma$ upper limit of 27mJy (see section 4.3) on emission
from this region. This is steeper than surface density profile of $\Sigma$$\propto r^{7/3}$ predicted in the models of \citet{2004ApJ...602L.133K}.  For a surface density profile like those of the delayed stirring models to be consistent with the observations, then the disc would have to be truncated at 24 AU.  Assuming the disc is being truncated by a planet at just inside 28AU and that the disc is currently being stirred, then this constraint on the secular timescale means a minimum mass for the perturber can be derived using Eq. 42 from \citet{1999ApJ...527..918W} which gives a minimum planet mass of 2.6M$_{\oplus}$ for a system age of 12 Myr.

\subsubsection{\label{sub:Delayed-Stirring-Model}Delayed Stirring Model}

Another possible interpretation of these observations is that the
resolved belt corresponds to a region of a more extended disc that
has recently formed Pluto sized objects. In these self-stirred models
of \citet{2005AJ...130..269K} the evolution of an extended planetesimal
belt is followed, allowing both the velocity and size distributions
to evolve self-consistently. A destructive collisional cascade is
only ignited in the disc when planets forming in the disc reach Pluto-size.
The timescale for the formation of these Pluto sized objects depends
strongly on their distance from the star and on the surface density
of the disc so the outer disc regions are stirred at later times.
We performed further modelling based on the prescriptions for the
evolution given in \citet{2008ARA&A..46..339W} and refined in Kennedy
\& Wyatt (2010) which provides an empirical fit to the self-stirred
models of \citet{2005AJ...130..269K}. This model consists of a
planetesimal belt with radii r$_{min}$ and r$_{max}$ with surface
density scaled to the MMSN divided into annuli at different radii
which have a suppressed dust production rate until the time at which
Pluto sized bodies have formed, stirring the annulus into a destructive
collisional cascade, increasing the production of small dust and causing
it to brighten. The dust emission is calculated assuming emission
from dust grains with a size distribution and composition consistent
with the SED and image modelling in previous sections and a ratio
of dust area to planetesimal mass appropriate for a collisional cascade
size distribution. To account for deviations from this theoretical
size distribution a scaling factor $\eta$ was applied to the surface
brightness of the whole disc. However, under these constraints from
the SED and with the stirring time set to the age of HD191089 (12Myr)
the model provided a poor fit to the observations. This is primarily
due to the shape of the surface density distribution produced.To produce
enough emission in the outer regions of the disc (i.e out to $\sim$90AU)
there has to be a region of gradually rising surface density (with
a power law form $\Sigma$$\propto r^{7/3}$ ) in the inner region
of the disc (see Kennedy \& Wyatt 2010 Fig. 2).  For reasons similar to those discussed in 5.1.1 this overproduces
emission at <28 AU (0.5''), where the 3$\sigma$ upper limit on
the flux in this region is 27mJy. Therefore the delayed stirring model
cannot fit the sharp inner edge to the disc.  However this can be achieved if there is an additional mechanism truncating the disc. For example a disc with an inner radius of 24 AU, surface density profile $\Sigma\propto r^{7/3}$ between
24 and 34 AU, and  $\Sigma\propto r^{-1.5}$ from 34 to 90 AU can fit the observations.
This gives a flux of 26mJy in the inner 28AU (0.5'') of the disc,
consistent with the 3$\sigma$ upper limit. We conclude that some mechanism must be truncating the disc, and that it is not possible to determine from these observations alone whether planet formation is ongoing in some or all of the outer disc.

\subsubsection{\label{sub:Protoplanetary-Disc-Remnant}Protoplanetary Disc Remnant}

A further possibility  is that this disc
is a remnant of the protoplanetary disc and the dust is confined in
a ring due to interactions with gas that remains in the system from
the protoplanetary phase (\citealp{2001ApJ...557..990T,2007ApJ...655..528B}). These models are too gas rich for HD191089 as a massive gas disc around a star of this age would be unusual
as most gas discs are thought to dissipate in <10Myr (e.g \citealp{2005ApJ...631.1180H}). Gas has been detected around 4 A stars known to have
debris discs ($\beta$ Pic: \citealp{2004A&A...413..681B}, 49 Ceti:
\citealp{2005MNRAS.359..663D}, HD141569: \citealp{2006A&A...453..163J},
HD32297: \citealp{2007ApJ...656L..97R}) and its effect on dust is debated;\citet{2009A&A...507.1503K} concluded that gas does not affect the dust distribution in $\beta$ Pic, though \citet{2007ApJ...655..528B} discussed how gas may be
undetected and still have an effect on the dust dynamics.
Deep gas observations towards HD191089 would be needed to rule out
this effect.

\subsection{Placing HD191089 into context: Comparison with F stars of the BPMG}

The BPMG (Beta Pictoris Moving Group) is a collection of young ($\sim$12Myr)
stars with common proper motion, indicating that they formed in the
same cluster which is now dispersing. Their velocities can be traced
back to this common point of origin. The age of the Moving Group
can be assessed using the M star members, and the group is then assumed
to be coeval. Table \ref{tab:BPMG Census} gives a census of the currently
known F star members of the $\beta$ Pic Moving Group. Membership
was taken from \citet{2009AJ....137.3632L}, who assessed the membership
of the moving group by examining stellar proper motions. By studying
the fraction of these stars that have infrared excesses we can place
HD191089 into context with its coeval moving group. Of the nine known
F star BPMG members, 6 have infrared excesses, 3 of which have resolved
debris discs: HD15115 (\citealp{2007ApJ...661L..85K}), HD181327 (\citealp{2006ApJ...650..414S,2008ApJ...689..539C})
and HD191089 (this work).

For the 6 stars with discs SED fitting was used to estimate radius and fractional luminosity, which are given in Table 2, and range from 10 to 150 AU, and from $3\times10^{-5}$ to $3\times10^{-3}$.  HD191089 is the second brightest and  has the second smallest inner radius. It is the most similar to HD181327 which is slightly brighter and larger. The similarity is also apparent in images.
HD181327 has been imaged in both scattered light using HST (\citealp{2006ApJ...650..414S})
and in mid-IR thermal emission (\citealp{2008ApJ...689..539C}. This
disc shows an extended icy {}``Kuiper belt'' at 86AU with a width
of 36AU (\citealp{2006ApJ...650..414S}), compared to a belt at 55AU
with a width of 50AU for HD191089. Both these discs have cleared inner
regions (HD181327 $\sim$68 AU, HD191089 $\sim$28AU)
and both show evidence for a brightness asymmetry between the two
sides of the disc in the mid-IR (HD181327: flux ratio of 1.4$\pm$0.1
between the Northern and Southern arms HD191089: flux ratio of 0.8$\pm0.12$
between the E and W sides of the disc). 

The other resolved disc HD15115 is a very large, highly asymmetric
disc, dubbed the {}``Blue Needle'' due to the elongation of the
West side of the disc to >550AU (East side $\sim$ 310AU) (\citealp{2007ApJ...661L..85K}).
Such an extreme asymmetry could be due to a stellar fly-by, and HIP12545
(also a member of the BPMG) is a possible candidate due to its near
on-sky position (\citealp{2007ApJ...661L..85K}). This asymmetry could also be caused by interaction with a nearby ISM cloud, that
could be 'sandblasting' and so stripping material from one side of
the disc (\citealp{2009ApJ...702..318D,2009ApJ...707.1098M}). There
are not many obvious similarities between HD191089 and HD15115: the unusual morphology of HD15115 appears to originate
through interaction with an external perturber rather than due to any planets in the system and it appears much more extended in images than HD191089. Vega's disc displays a similarly extended shape to HD15115, with a high population
of small grains in the process of being blown out of the system (\citealp{2005ApJ...628..487S}). However,
this system can still be explained in the context of a steady state
collisional cascade when the difference in stellar luminosity between
the pole and equator due to Vega's fast rotation is taken into account
(\citealt{2010ApJ...708.1728M}). 

The remaining three stars are all binaries, and one of these, HD199143,
may show some evidence for an infrared excess due to a circumsecondary
disc (\citealp{2001ApJ...561L.111J}), but this may simply be due
to uncertainties in the photospheric modelling of the binary companion
(\citealp{2000A&A...363L..25V,2002A&A...394..219C,2004A&A...414..175K}).
The general lack of excess seen for the F stars of the BPMG that are
also binaries could be due to perturbations from the secondary truncating
any possible discs. HD199143 has a seperation of $\sim$53-105AU (\citealp{2001ApJ...561L.111J,2002A&A...394..219C}) and there is observational evidence that binary companions in the range 50-100AU are expected to disrupt discs (\citealp{1994ApJ...429L..29J}. The other two binaries have quite wide
predicted separations ( HD14082: $\sim$530AU \citealp{2003ApJ...599..342S},
HD29391: $\sim$800AU \citealp{2006AJ....131.1730F}), and unless
on very eccentric orbits the secondaries are unlikely to preclude
the presence of discs around the primaries in these systems. It is possible that the binary has evolved or that there is something
about the formation mechanism of binary systems that suppresses formation
or detectability of a debris disc, but the small number statistics
for this sample means that this is still inconclusive.

\begin{table*}
\caption{\label{tab:BPMG Census} Census of F star members of the BPMG. The
dust radius is calculated by fitting a black body to the 24$\umu$m
and 70$\umu$m fluxes, or from previous literature black body fits
to the SED.}

\begin{tabular}{>{\centering}p{2cm}>{\centering}p{1.5cm}>{\centering}p{1.5cm}>{\centering}p{1.5cm}>{\centering}p{1.5cm}c>{\centering}p{2cm}>{\centering}p{1cm}>{\centering}p{2.2cm}}
\toprule {\footnotesize Star }  & {\footnotesize Spectral Type}  & {\footnotesize Mass ($M_{\odot}$) }  & {\footnotesize Distance (pc) }  & {\footnotesize Luminosity ($L_{\odot}$) }  & {\footnotesize $\frac{L_{Dust}}{L_{*}}$}  & {\footnotesize Blackbody Dust Radius {[}Resolved{]} (AU) }  & {\footnotesize Refs }  & {\footnotesize Notes}\tabularnewline
\midrule {\footnotesize HD203 }  & {\footnotesize F3V }  & {\footnotesize 1.3 }  & {\footnotesize 39.1 }  & {\footnotesize $2.7$ }  & {\footnotesize $1\times10^{-4}$}  & {\footnotesize 10 }  & {\footnotesize 1, 2,3,4}  & {\footnotesize 1}\tabularnewline
{\footnotesize HD 14082A }  & {\footnotesize F5V }  & {\footnotesize 1.23 }  & {\footnotesize 39.42 }  & {\footnotesize $1.88$ }  & {\footnotesize N/A}  & {\footnotesize N/A }  & {\footnotesize 1,2,3,4 }  & {\footnotesize 2 (Binary)}\tabularnewline
{\footnotesize HD 15115 }  & {\footnotesize F2V }  & {\footnotesize 1.39 }  & {\footnotesize 44.8 }  & {\footnotesize 2.94 }  & {\footnotesize $4.9\times10^{-4}$}  & {\footnotesize 34 {[}\textasciitilde{}300-500{]} }  & {\footnotesize 5,6,7}  & {\footnotesize 3 (Resolved disc)}\tabularnewline
{\footnotesize HD 29391 }  & {\footnotesize F0V }  & {\footnotesize 1.27 }  & {\footnotesize 29.9 }  & {\footnotesize 5.85 }  & {\footnotesize N/A}  & {\footnotesize N/A }  & {\footnotesize 1}  & {\footnotesize 4 (Binary)}\tabularnewline
{\footnotesize HD 35850 }  & {\footnotesize F7V }  & {\footnotesize 1.19 }  & {\footnotesize 26.8 }  & {\footnotesize 1.78 }  & {\footnotesize $3\times10^{-5}$}  & {\footnotesize 150 }  & {\footnotesize 1,8,9,10,11 }  & {\footnotesize 5}\tabularnewline
{\footnotesize HD 164249 }  & {\footnotesize F5V }  & {\footnotesize 1.32 }  & {\footnotesize 46.8 }  & {\footnotesize 2.64 }  & {\footnotesize $5.9\times10^{-4}$}  & {\footnotesize 20 }  & {\footnotesize 1}  & {\footnotesize 6}\tabularnewline
{\footnotesize HD 181327 }  & {\footnotesize F5/6V }  & {\footnotesize 1.3 }  & {\footnotesize 51 }  & {\footnotesize 3.1 }  & {\footnotesize 3$\times10^{-3}$}  & {\footnotesize 20 {[}86{]} }  & {\footnotesize 7,1,6,12,13 }  & {\footnotesize 7 (Resolved disc)}\tabularnewline
{\footnotesize HD 191089 }  & {\footnotesize F5V }  & {\footnotesize 1.4 }  & {\footnotesize 54 }  & {\footnotesize 3.2 }  & {\footnotesize 1.4$\times10^{-3}$}  & {\footnotesize 15 {[}55{]} }  & {\footnotesize 1,7,8,14}  & \tabularnewline
{\footnotesize HD 199143 }  & {\footnotesize F8V }  & {\footnotesize 1.2 }  & {\footnotesize 48 }  & {\footnotesize 2.45 }  & {\footnotesize ?}  & {\footnotesize ? }  & {\footnotesize 15,16,17,18 }  & {\footnotesize 8}\tabularnewline
\bottomrule  &  &  &  &  &  &  &  & \tabularnewline
\end{tabular}

\begin{minipage}[t][1\textheight]{0.15\textwidth}%
\vspace*{0.7cm}

\textbf{References:}

1. \citealp{2008ApJ...681.1484R}

2. \citealp{2008ApJS..179..423C}

3. \citealp{2009AJ....137.3632L}

4. \citealp{2009ApJS..181..197C}

5. \citealp{2007ApJ...661L..85K}

6. \citealp{2006ApJ...644..525M}

7.\citealp{2006ApJS..166..351C}

8. \citealp{2009A&A...497..409R}

9. \citealp{1997A&A...323L..49P}%
\end{minipage}~~~~~%
\begin{minipage}[t]{0.25\textwidth}%
\vspace*{0.7cm}

10. \citealp{2001ApJ...555..932S}

11. \citealp{2006ApJ...639.1138S}

12. \citealp{2008ApJ...689..539C}

13. \citealp{1993yCat.2156....0M}

14. \citealp{2006ApJS..166..351C}

15.\citealp{2001ApJ...561L.111J}

16.\citealp{2004A&A...414..175K}

17. \citealp{2000A&A...363L..25V}

18. \citealp{2002A&A...394..219C}%
\end{minipage}~~~~~~~%
\begin{minipage}[t]{0.45\textwidth}%
\vspace*{0.7cm}
 \textbf{Notes:}

1. Strong Spitzer $24\umu m$ excess ($F_{24}=60F_{*24}$) \citealp{2009ApJS..181..197C}

2. Binary companion HD 14082B (G5V, 530Au separation). \citealp{2003ApJ...599..342S}

3. Resolved Asymmetric Debris disc {}``Blue Needle''. Possibly perturbed
by HIP12545 flyby. \citealp{2007ApJ...661L..85K}

4. Binary companion (\textasciitilde{}800AU). Blended with Spitzer.

5. Dust based on Spitzer 70$\umu m$ excess. \citealp{2006ApJ...639.1138S}

6. Dust based on Spitzer $24\umu m$ and 70$\umu m$ excess. \citealp{2008ApJ...681.1484R}

7. Resolved disc at \textasciitilde{}86AU (\textasciitilde{}20AU width).\citealp{2006ApJ...650..414S,2008ApJ...689..539C}

8. Binary companion (M2, 86AU \citealp{2001ApJ...561L.111J,2004A&A...414..175K}).
Possible circumsecondary disc (\citealp{2000A&A...363L..25V,2002A&A...394..219C,2004A&A...414..175K}).%
\end{minipage}
\end{table*}

\section{Conclusions}

We have presented Qa band Gemini T-ReCs imaging of the 12 Myr old
F star HD191089. The emission at 18.3$\umu$m is shown to be significantly
extended compared with a point source. This image represents the first
resolution of dust emission around this star. These observations confirm
the interpretation of the SED as a debris disc with a single component
resolved at 28-90 AU. There is little emission from the inner 28AU
of the disc, indicating the presence of an inner cavity.

The disc also shows a tentative brightness asymmetry of $\frac{F_{W}}{F_{E}}=0.80\pm0.12$
with a significance of 1.8$\sigma$. This asymmetry is consistent
with a scenario in which an perturbing body on an eccentric orbit
imposes a forced eccentricity of 0.12 on the planetesimal belt through
secular perturbations, causing the centre of the ring to become offset
from the star which replicates the brightness asymmetry observed.
This interpretation can be tested
by further observations seeking to confirm the brightness asymmetry
and measure the predicted offset of the star from the centre of the
ring.

 To predict the surface brightness of the  HD191089 disc in scattered light we used $S=\frac{F\tau\omega}{4\pi\phi^{2}}$ (\citealp{1999ApJ...525L..53W}) where S is the surface brightness (mJy/arcsec$^{2}$), F is the stellar flux (mJy), $\tau$ is the optical depth, $\omega$ is the albedo and $\phi$ is the angular distance of the disc from the star. Using our
best fit model parameters at NICMOS wavelengths ($\lambda_c$=1.1, $\Delta\lambda$=0.59)  we predict
a surface brightness of 0.024-0.19mJy/arcsec$^{2}$ assuming isotropic
scattering and an albedo of 0.1-0.8 the range predicted for Kuiper
belt objects (\citealp{2008ssbn.book..161S}). Although this is detectable
with ACS and NICMOS resolving the disc would be difficult as the peak
surface brightness is expected at a radius of 1'', which will be
close to the inner working edge of the coronograph for NICMOS observations,
and inside it for ACS.

Three models for the central hole were considered. The interpretation
that fits most neatly with the model for the brightness asymmetry
is that the inner region has been cleared of planetisimals by a planetary
system. In this case high resolution images of the system could resemble
Fomalhaut, including a planet close to the inner edge of the belt.
We found that the sharpness of the inner edge was incompatible with
the delayed stirring models (\citealp{2004ApJ...602L.133K}) in which
the disc is in the process of forming Pluto sized objects causing
the disc to brighten at the radii where they have just formed, although it is possible that planet formation of the type envisioned by these models is ongoing as long as some other process is truncating the disc at 25AU. A third
possibility is that the planetesimal belt is somehow confined due
to dynamical interactions with gas in the system. However, there are
no observational constraints on the presence of gas in the HD191089
system. When compared to other members of the BPMG, the most obvious
similarities in terms of fractional luminosity and radius inferred
from blackbody fitting are with HD181327, which has also been resolved
at mid-IR wavelengths (\citealp{2008ApJ...689..539C}) and shows evidence
for an inner clearing and asymmetry similar to those seen in HD191089.

\section{Acknowledgments}

L. J. C. is grateful for the support of an STFC studentship. R. S.
is grateful for the support of the STFC. The authors would like to thank Christine
Chen for providing the Spitzer IRS spectrum of HD191089. The authors would also like to thank Tim Gledhill for his useful comments that much imrpoved this paper.  Based on
observations obtained at the Gemini Observatory, which is operated
by the Association of Universities for Research in Astronomy, INC.,
under a cooperative agreement with the NSF on behalf of the Gemini
partnership: The National Science Foundation (United States), the
Science and Technology Facilities Council (United Kingdom), the National
Research Council (Canada), CONICYT (Chile), the Australian Research
Council (Australia), ministirio da Ciencia e Tecnologia (Brazil),
and SECYT (Argentina).

\bibliographystyle{mn2e}
\bibliography{references}
 \label{lastpage} 
\end{document}